# Two-photon interference of polarization-entangled photons in a Franson interferometer

Heonoh Kim[1], Sang Min Lee[2], Osung Kwon[3], and Han Seb Moon[1,*]

[1]Department of Physics, Pusan National University, Geumjeong-Gu, Busan 46241, South Korea
[2]Korea Research Institute of Standards and Science, Daejeon 34113, South Korea
[3]National Security Research Institute, Daejeon 34044, South Korea
*Corresponding author: hsmoon@pusan.ac.kr

## Abstract

We present two-photon interference experiments with polarization-entangled photon pairs in a polarization-based Franson-type interferometer. Although the two photons do not meet at a common beamsplitter, a phase-insensitive Hong-Ou-Mandel type two-photon interference peak and dip fringes are observed, resulting from the two-photon interference effect between two indistinguishable two-photon probability amplitudes leading to a coincidence detection. A spatial quantum beating fringe is also measured for nondegenerate photon pairs in the same interferometer, although the two-photon states have no frequency entanglement. When unentangled polarization-correlated photons are used as an input state, the polarization entanglement is successfully recovered through the interferometer via delayed compensation.

## Introduction

The Hong-Ou-Mandel (HOM) interference[1] is a very important two-photon quantum interference phenomenon in modern quantum mechanics and experimental quantum optics[2,3]. In the field of quantum information science and technology, observation of the HOM effect is essential not only to certify the indistinguishability of distinct photon sources, but also to prepare an entangled state from independent single photons[4-17]. The HOM effect has been considered as two-photon spatial bunching at one of two output ports when two identical photons impinge on a beamsplitter (BS) simultaneously. By completely superposing the two incoming photons at the

BS, it is possible to cause the coincidence detection events to vanish at the two BS output ports. Quantum mechanical interpretation of the HOM effect relies on the interference of two alternative probability amplitudes leading to coincidence detection by two single-photon detectors located at the BS output ports. Typically, Feynman-like path diagrams for the two possible processes at the BS have been used to explain two-photon interference effects, especially for HOM-type experiments[18,19]. This approach can also be applied to explicate various quantum interference experiments employing correlated photons and using Mach-Zehnder and Michelson interferometers[20,21].

To date, HOM-type two-photon interference fringes have only been observed when the two input photons are mixed at a common BS, regardless of whether the two photons meet simultaneously at the BS[1,18,21]. This invites the question of whether the HOM-type fringe pattern can be observed without using a common BS. The answer to this question is closely dependent on whether the two photons must be mixed in a common BS to produce an interference fringe. To verify observation of the HOM-type interference fringe without using a BS, we newly designed a modified Franson interferometer consisting of two local interferometers having two unbalanced arms[22]. In the Franson interferometer, two individual photons do not meet at a common BS nor are they mixed against other optical interferometers. Note that observation of the HOM interference fringe without mixing two photons via a common BS is very important as regards consistent understanding of the various kinds of quantum interference phenomena involving correlated multiphoton states.

In this paper, we experimentally demonstrate the observation of HOM-type two-photon interference fringes in the polarization-based Franson-type interferometer, in which there is no BS for two-photon crossings. For the input state, we employ polarization-entangled photon pairs generated via spontaneous parametric down-conversion (SPDC) using type-II noncollinear quasi-phase matching (QPM) in a periodically poled $KTiOPO_4$ (PPKTP) crystal[23]. The interferometer is appropriately designed using polarizing beamsplitters (PBSs), quarter-wave plates, and a double-sided mirror within a single-rail configuration as one of the interferometer arms. As a result, the phase-insensitive HOM interference peak and dip fringes are successfully measured through post-selection of the polarization states. HOM fringes with nondegenerate photon pairs are also observed in the same interferometric scheme. Moreover, the HOM fringes are observed through delayed compensation at the Franson-type interferometer despite the fact that two-photon states of which polarization entanglement was degraded are injected into the interferometer.

## Results

To observe the HOM interference fringes, a time delay is introduced between two photons incident on the BS. There are two methods of achieving this delay. In one approach, the BS itself is moved from the symmetric position within the interferometer; this method was used in the original HOM scheme[1]. In this case, no time delay is introduced between the two transmission amplitudes (*TT*) for the two input photons when the BS is moved, whereas an asymmetric time delay is introduced between the two reflection amplitudes (*RR*). In the second method, an optical delay is introduced at only one of the interferometer arms[24]. For the former method, if we ignore whether the paths of the two incoming photons towards the two detectors are exchanged, for the *TT* amplitude, the BS can be replaced by a double-sided mirror, which yields the *RR* amplitudes necessary to observe the HOM fringe without mixing the two incoming photons. To achieve this mechanism, we have designed a novel polarization-based Franson-type two-photon interferometer.

**Experimental setup.** Figure 1 shows the experimental setup used to observe the HOM-type interference fringes without using BS for mixing two incoming photon pairs. Our experimental setup is composed of two interferometric parts: the two-photon polarization state preparation component, which acts as a PBS-based birefringence compensator[23,25], and the PBS-based Franson-type two-photon interferometer. The two-photon polarization state is generated in a PPKTP crystal through a type-II noncollinear QPM-SPDC process[23], which is described in mixed-state form as $\rho_{\text{PPKTP}} = 1/2(|\psi\rangle\langle\psi| + |\phi\rangle\langle\phi|)$, where $|\psi\rangle = |H,\omega_1\rangle_a |V,\omega_2\rangle_b$ and $|\phi\rangle = |V,\omega_2\rangle_a |H,\omega_1\rangle_b$. Here, *H* (*V*) denotes the horizontal (vertical) polarization, $\omega_i$ is the photon angular frequency, and the subscripts *a* and *b* represent the two spatial modes for the two photons. Using the half-wave plate placed in one of the arms, both two-photon polarization states are transformed to $|H,\omega_1\rangle_a |H,\omega_2\rangle_b$ and $|V,\omega_2\rangle_a |V,\omega_1\rangle_b$. When there is no path-length difference from the PPKTP to PBS0, i.e., $\Delta x_0 = 0$ (Fig. 1), the polarization-entangled photon pairs are generated at the two output ports of the PBS0[23,25]. Then, the two-photon state is injected into each PBS of the Franson interferometer.

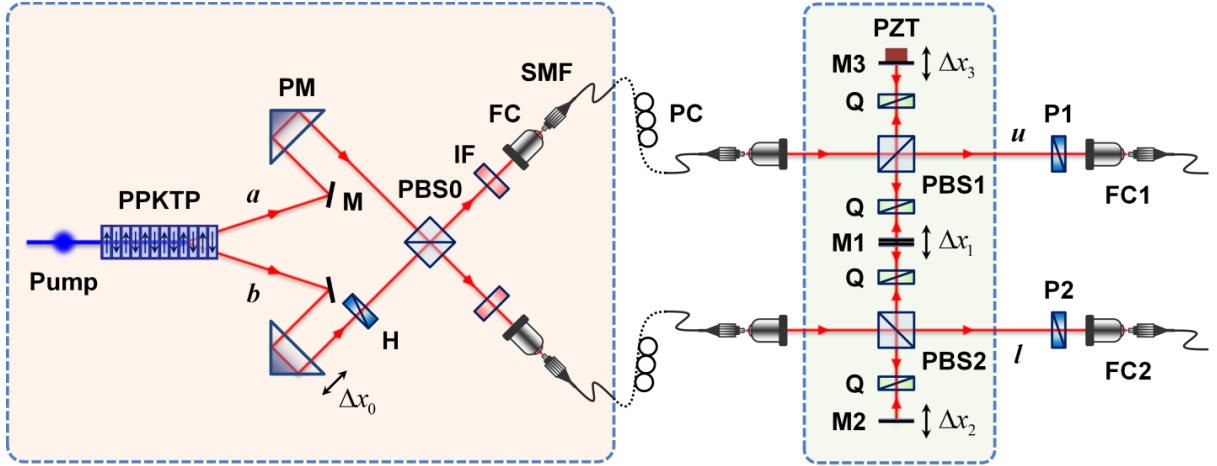

**Figure 1. Experimental setup to observe two-photon interference effect.** Pump, continuous-wave (cw) single-mode diode laser (406.2 nm, 5 mW); PM, prism mirror; M, mirror; H, half-wave plate; PBS, polarizing beamsplitter; IF, interference filter; FC, fibre coupler; SMF, single-mode fibre; PC, polarization controller; Q, quarter-wave plate; PZT, piezoelectric transducer; P1 and P2, linear polarizers. In particular, we use a double-sided mirror (M1) for simultaneous adjustment of the path-length differences between the two long interferometer arms in the opposite direction, while the relative phase differences between them always remain at zero.

The two-photon state at the interferometer output ports is described in the form

$$|\Phi\rangle_{out} = \frac{1}{\sqrt{2}} \left( |T, \omega_2\rangle_u |T, \omega_1\rangle_l + e^{i\beta} |R, \omega_2\rangle_u |R, \omega_1\rangle_l \right), \quad (1)$$

where the subscripts denote the two photons passing through the upper ($u$) and lower ($l$) interferometers, respectively, and the phase factor $\beta$ arises from the relative phase difference between the two long interferometer arms. $T$ and $R$ represent the transmission and roundtrip amplitudes for the photons in the PBS-based interferometer, respectively, which correspond to the short and long paths in the original Franson scheme[22]. Here, the phase factor $\beta = \phi_1 + \phi_2$ is related to the optical path-length difference, where $\phi_1 = k\Delta L_u$, $\phi_2 = k\Delta L_l$, here, $k$ is a wave vector and $\Delta L_u$ and $\Delta L_l$ are the optical path lengths of the two long (upper and lower) interferometer arms. In the experiment, when the path-length difference $\Delta x_1$ is adjusted by moving M1, the optical paths of the two long interferometer arms are simultaneously changed in the opposite direction, $\pm \Delta L_u = \mp \Delta L_l$, so the overall phase difference, $\delta \beta = \delta(\phi_1 - \phi_2)$, between the two long interferometer arms always remain at zero. That is, an additional phase difference cannot be introduced by changing the M1 position. The necessary conditions for observation of two-photon interference are: (i) The path-length difference between the two long arms must be shorter than the single-photon coherence length; and (ii) The path-length difference between the two two-photon amplitudes in Eq. (1) must be shorter than the two-photon coherence length. Furthermore, the polarization information should be erased using polarization analyzers placed before single-mode fibre couplers (FCs).

Our interferometer scheme is modified from the original Franson setup such that the two PBSs are employed to remove two unwanted two-photon amplitudes corresponding to two unbalanced short-long and long-short paths[26,27]. In particular, we employ a double-sided mirror M1, as shown in Fig. 1, in a single-rail configuration to adjust the path-length differences between the long interferometer arms. This modification is also required in order to improve the experimental facility for observation of the phase-insensitive HOM-type interference fringe patterns because the path-length differences between the two long interferometer arms must be adjusted simultaneously without introducing an additional relative phase difference. In our experiment, the path-length difference with constant phase in the long arms induced by movement of M1 is essential for observation of the HOM interference fringe in the interferometer. If we ignore the propagation time difference between the $|T\rangle_u |T\rangle_l$ and $|R(\pm \Delta x_1)\rangle_u |R(\mp \Delta x_1)\rangle_l$ amplitudes, the M1 movement ($\pm \Delta x_1$) has the same function as the BS movement used in the original HOM scheme. Two mirrors M1 and M2 are mounted on motorized translation stages while M3 is mounted on a PZT actuator for observation of the phase-sensitive oscillatory interference fringes. Two linear polarizers (P1 and P2) are placed at the interferometer output ports to erase the polarization information on the two-photon polarization states. Finally, two-photon interference fringes are observed via coincidence counting, which is performed by two single-photon detectors (SPCM-AQRH-FC, Excelitas).

**HOM interference of frequency-degenerate polarization-entangled photons.**

Figure 2 shows the experimental results. Phase-insensitive HOM-type two-photon interference fringes are clearly observed as a function of $\Delta x_1$ when a polarization-entangled state with degenerate photon pairs is employed as an input state, as shown in Fig. 2(a). The filled squares (circles) represent the HOM fringe peak (dip) as measured for a polarization analyzer angle combination of $\theta_{P1} = +45°$ and $\theta_{P2} = +45°$ ($\theta_{P2} = -45°$). The solid lines represent the theoretical fit to the experimental data. The measured coincidence counting rate at the plateau is approximately 2,000 Hz. The observed peak and dip fringe visibilities are found to be 0.92 ± 0.02 and 0.93 ± 0.02, respectively.

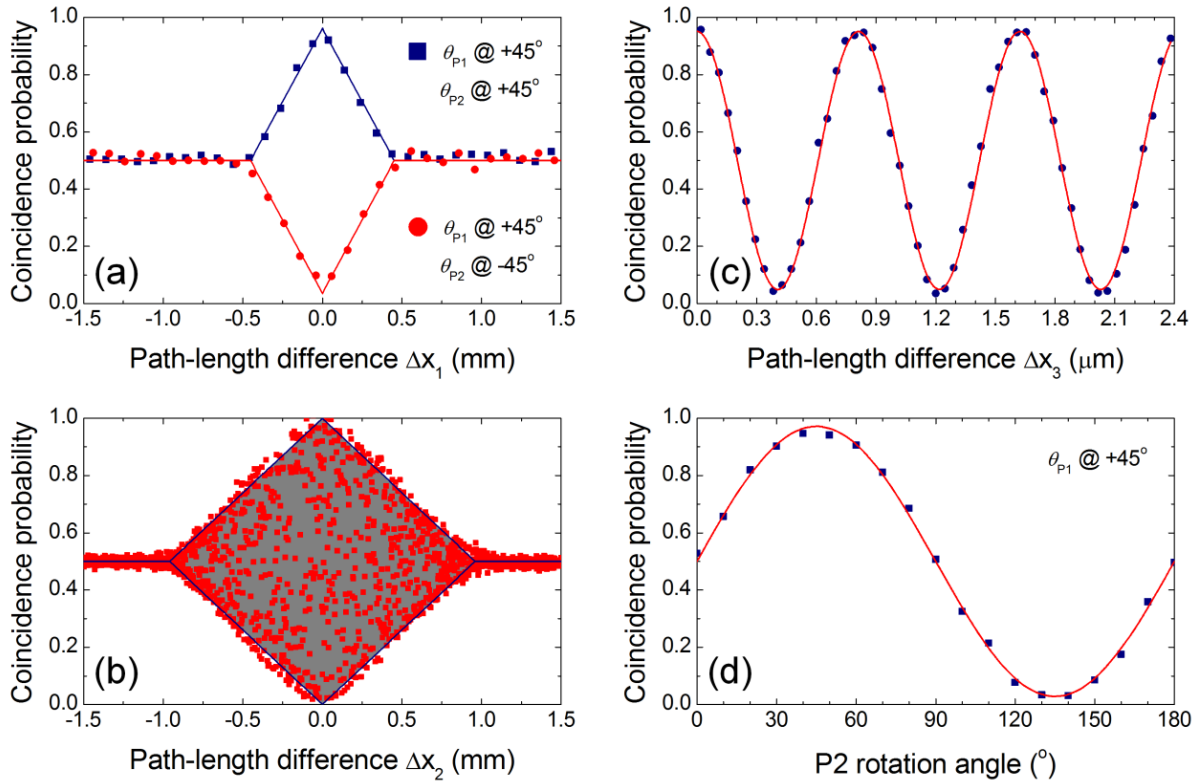

**Figure 2. Experimental results.** (**a**) Measured HOM peak and dip fringes as a function of $\Delta x_1$ for degenerate photon pairs when $\Delta x_0 = 0$. (**b**) Coincidence probability as a function of $\Delta x_2$ for $\Delta x_1 = 0$. (**c**) Phase resolved fringe of (**b**) measured by varying $\Delta x_3$ for $\Delta x_2 \approx 0$. (**d**) Polarization correlation measured as a function of P2 rotation angle, with P1 fixed at $+45°$.

The HOM fringe exhibits a triangular shape over the range of the single-photon coherence length[23,28]. This intrinsic fringe shape for the photon-pair source used in our experiment is related to the natural rectangular-shaped two-photon wave function in type-II SPDC process[29,30]. The coincidence counting probability is expressed as a function of the path-length differences between the two long interferometer arms in the form

$$P = \frac{1}{2}\left[1 \pm V f(\Delta x_1)\cos\left(\Delta\omega \frac{\Delta x_1}{c}\right)\cos\left(\frac{2\pi}{\lambda}|\Delta x_2 - \Delta x_3|\right)\right], \quad (2)$$

where $V$ is the fringe visibility, $c$ is the speed of light in vacuum, $\lambda$ is the single-photon center wavelength, $f(\Delta x_1)$ is an envelope function corresponding to the spectral property of the detected single-photon wavepackets, and $\Delta\omega$ is related to the beat frequency between the two photons[20,21,23,31,32]. $\Delta x_2$ ($\Delta x_3$) is the change in position of M2 (M3), as shown in Fig. 1. A coincidence peak and dip can be observed where there is no path-length difference between the two long arms ($\Delta x_1 = 0$ and $|\Delta x_2 - \Delta x_3| = 0$). Meanwhile, when a path-length difference is introduced by moving M2 with M1 fixed ($\Delta x_1 = 0$), we observe a phase-sensitive oscillatory interference fringe within the peak/dip envelope, as shown in Fig. 2(b). In this case the fringe width is twice that in Fig. 2(a), because only one of the paths is adjusted. The phase-resolved fringe of Fig. 2(b) is measured by varying $\Delta x_3$ (the M3 position, see Fig. 1) when there is no path-length difference in the long arms of the interferometer, as shown in Fig. 2(c). The fringe visibility is found to be 0.91 ± 0.01. Additionally, we verify the polarization correlation between the two output photons by rotating P2 with P1 fixed at $+45°$, for $\Delta x_i = 0$ ($i = 1, 2, 3$). In this experiment, the measured coincidence probability should obey the relation $P \propto 1 + V\cos[2(\theta_{P2} - \theta_{P1})]$ for the $|\Phi\rangle = 1/\sqrt{2}(|H\rangle_u|H\rangle_l + |V\rangle_u|V\rangle_l)$ state. In Fig. 2(d), the fringe visibility is 0.94 ± 0.01, which clearly shows the polarization entanglement of the detected photon pairs.

## HOM interference of frequency-nondegenerate polarization-entangled photons.

If the input photons have different center wavelengths for nondegenerate QPM-SPDC conditions[23], the spatial beating fringe can also be observed in our interferometer; this is despite the fact that the two incoming photons are not in a frequency-entangled state in the two interferometer arms, as shown in Eq. (1). This feature implies that the spatial beating fringe does not necessarily involve the frequency-entangled state which is different behavior to that examined in previously reported works[31,32]. The key aspect of the two-photon quantum interference effect is not the initial state at the interferometer input port but, rather, the final state at the detection stage, in which all two-photon amplitude distinguishability is completely eliminated. Figure 3 shows the spatial beating fringe measured as a function of $\Delta x_1$. The peak and dip fringe visibilities are 0.92 ± 0.02 and 0.91 ± 0.03, respectively. The frequency difference estimated from the oscillation period is 0.69 ± 0.01 THz, which corresponds to 1.52 nm of the center-wavelength difference between two single photons.

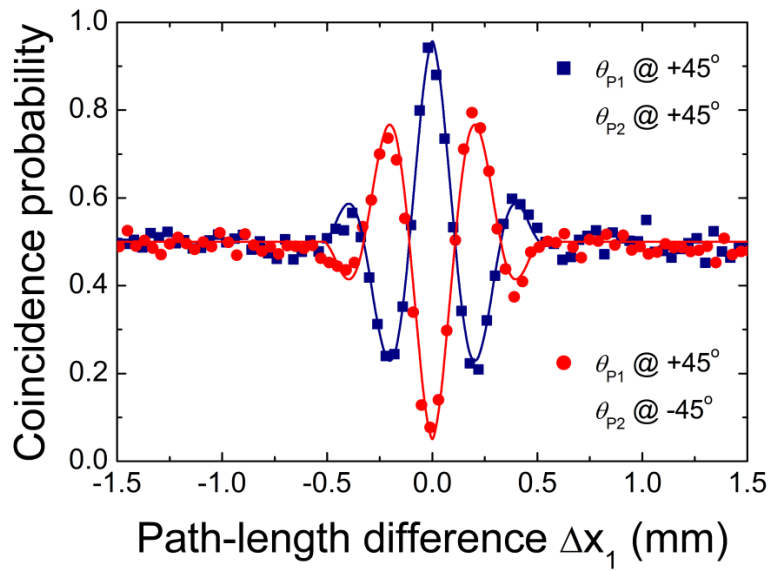

**Figure 3**. Measured HOM peak and dip fringes as a function of $\Delta x_1$ for nondegenerate photon pairs when $\Delta x_0 = 0$.

**HOM interference of frequency-degenerate polarization-unentangled photons.**
Finally, we show that the HOM interference fringes can be observed despite the fact that two-photon states of which polarization entanglement was degraded are injected into the interferometer. To achieve this experimental condition, we adjust $\Delta x_0$ (shown in Fig. 1) to exceed the single-photon coherence length. In this case, if we ignore the group delay between the two polarization components in the SPDC process, the single-photon polarization component $|H\rangle_a$ ($|V\rangle_a$) is always preceded by $|H\rangle_b$ ($|V\rangle_b$) by the magnitude of $\Delta x_0$ at the PBS0 output ports. However, our interferometer can act as a postponed compensator here, similar to the Pittman's scheme for polarization entanglement recovery[18], except that the input state before PBS0 has two two-photon polarization states $|H\rangle_a |H(\Delta x_0)\rangle_b$ and $|V\rangle_a |V(\Delta x_0)\rangle_b$. Figure 4 shows the HOM fringes measured as a function of $\Delta x_1$ for $\Delta x_0 = +1$ mm. As predicted, interference peak and dip positions are observed when M1 is moved toward the lower arms by the same degree as $\Delta x_0$; this is because the overall path-length difference between the two long-interferometer arms must be $2\Delta x_0$. This result demonstrates that the polarization entanglement is successfully recovered through the Franson interferometer by means of delayed compensation.

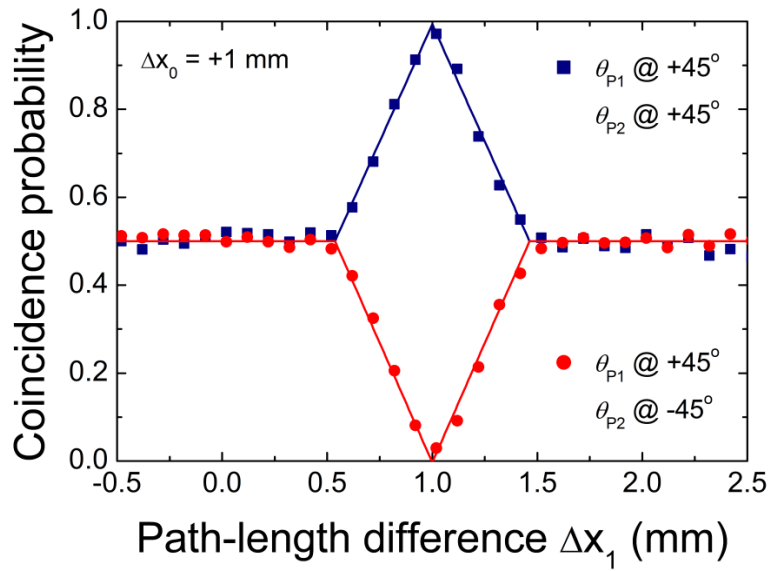

**Figure 4**. Measured HOM peak and dip fringes as a function of $\Delta x_1$ for degenerate photon pairs when $\Delta x_0 = +1$ mm.

## Conclusion

In summary, we have experimentally demonstrated the observation of HOM-type interference fringes without the use of a BS to mix two photons. This experiment was performed using a newly designed polarization-based Franson-type interferometer involving polarization-entangled photon pairs. Simultaneous adjustment of path lengths in one of the interferometer arms in the opposite direction facilitated introduction of a path-length difference only, without introduction of an additional relative phase difference between the interferometer arms. Although two photons were not mixed at the BS, which is contrary to the generally employed approach, HOM-type two-photon interference fringes were successfully observed. These interference fringes were obtained because of the quantum interference between the two alternative two-photon amplitudes related to the coincidence detection. When frequency-nondegenerate polarization-entangled photons were employed, spatial quantum beating fringes were observed, although the two photons were not frequency-entangled state within the interferometer arms. Finally, we also demonstrated successful recovery of the polarization entanglement through delayed compensation at the PBS-based Franson interferometer. The present results will facilitate a consistent understanding of various kinds of quantum interference phenomena involving correlated multiphoton states. Moreover, our scheme and results can aid development of new methods applicable to experimental quantum information science and technology.

## References

1. Hong, C. K., Ou, Z. Y. & Mandel, L. Measurement of subpicosecond time intervals between two photons by interference. *Phys. Rev. Lett.* **59**, 2044-2046 (1987).
2. Yuan, Z. S. *et al*. Entangled photons and quantum communication. *Phys. Rep*. **497**, 1-40 (2010).
3. Pan, J. W. *et al*. Multiphoton entanglement and interferometry. *Rev. Mod. Phys*. **84**, 777-838 (2012).
4. Pan, J. W., Bouwmeester, D., Weinfurter, H., & Zeilinger, A. Experimental entanglement swapping: entangling photons that never interacted. *Phys. Rev. Lett*. **80**, 3891-3894 (1998).

5. Lettow, R. *et al*. Quantum interference of tunably indistinguishable photons from remote organic molecules. *Phys. Rev. Lett*. **104**, 123605 (2010).

6. Stevenson, R. M. *et al*. Indistinguishable entangled photons generated by a light-emitting diode. *Phys. Rev. Lett*. **108**, 040503 (2012).

7. He, Y. *et al*. Indistinguishable tunable single photons emitted by spin-flip Raman transitions in InGaAs quantum dots. *Phys. Rev. Lett*. **111**, 237403 (2013).

8. Bocquillon, E. *et al*. Coherence and indistinguishability of single electrons emitted by independent sources. *Science* **339**, 1054-1057 (2013).

9. Lang, C. *et al*. Correlations, indistinguishability and entanglement in Hong-Ou-Mandel experiments at microwave frequencies. *Nat. Physics* **9**, 345-348 (2013).

10. Jin, H. *et al*. On-chip generation and manipulation of entangled photons based on reconfigurable lithium-niobate waveguide circuits. *Phys. Rev. Lett*. **113**, 103601 (2014).

11. Sipahigil, A. *et al*. Indistinguishable photons from separated silicon-vacancy centers in diamond. *Phys. Rev. Lett*. **113**, 113602 (2014).

12. Fakonas, J. S., Lee, H., Kelaita, Y. A. & Atwater, H. A. Two-plasmon quantum interference. *Nat. Photonics* **8**, 317-320 (2014).

13. Northup, T. E. & Blatt, R. Quantum information transfer using photons. *Nat. Photonics* **8**, 356-363 (2014).

14. Lopes, R. *et al*. Atomic Hong–Ou–Mandel experiment. *Nature* **520**, 66-68 (2015).

15. Toyoda, K., Hiji, R., Noguchi, A. & Urabe, S. Hong-Ou-Mandel interference of two phonons in trapped ions. *Nature* **527**, 74-77 (2015).

16. Xiong, C. *et al*. Active temporal multiplexing of indistinguishable heralded single photons. *Nat. Commun*. **7**, 10853 (2016).

17. Autebert, C. *et al*. Integrated AlGaAs source of highly indistinguishable and energy-time entangled photons. *Optica* **3**, 143-146 (2016).

18. Pittman, T. B. *et al*. Can two-photon interference be considered the interference of two photons?. *Phys. Rev. Lett*. **77**, 1917-1920 (1996). See also Pittman, T. B. Ph. D. Thesis, Univ. Maryland Baltimore County, 1996, Chapter 4.

19. Feynman, R. P., Leighton, R. B. & Sands, M. The Feynman Lectures on Physics, Vol. III (Addison-Wesley Publishing Co., Reading, MA, 1965).

20. Kim, H., Lee, S. M. & Moon, H. S. Generalized quantum interference of correlated photon pairs. *Sci. Rep*. **5**, 9931 (2015).

21. Kim, H., Lee, S. M. & Moon, H. S. Two-photon interference of temporally separated photons. *Sci. Rep*. **6**, 34805 (2016).

22. Franson, J. D. Bell inequality for position and time. *Phys. Rev. Lett.* **62**, 2205-2208 (1989).

23. Lee, S. M., Kim, H., Cha, M. & Moon, H. S. Polarization-entangled photon-pair source obtained via type-II non-collinear SPDC process with PPKTP crystal. *Opt. Express* **24**, 2941-2953 (2016).

24. Rarity J. G. & Tapster, P. R. Fourth-order interference in parametric downconversion. *J. Opt. Soc. Am. B* **6**, 1221-1226 (1989).

25. Kim, Y. H., Kulik, S. P., Chekhova, M. V., Grice, W. P. & Shih, Y. H. Experimental entanglement concentration and universal Bell-state synthesizer. *Phys. Rev. A* **67**, 010301 (2003).

26. Strekalov, D. V., Pittman, T. B., Sergienko, A. V., Shih, Y. H. & Kwiat, P. G. Postselection-free energy-time entanglement. *Phys. Rev. A* **54**, R1–R4 (1996).

27. Kwon, O., Park, K. K., Ra, Y. S., Kim, Y. S. & Kim, Y. H. Time-bin entangled photon pairs from spontaneous parametric down-conversion pumped by a cw multi-mode diode laser. *Opt. Express*, **21**, 25492-25500 (2013).

28. Fedrizzi, A. *et al*. Anti-symmetrization reveals hidden entanglement. *New J. Phys*. **11**, 103052 (2009).

29. Rubin, M. H., Klyshko, D. N., Shih, Y. H. & Sergienko, A. V. Theory of two-photon entanglement in type-II optical parametric down-conversion. *Phys. Rev. A* **50**, 5122-5133 (1994).

30. Sergienko, A. V., Shih, Y. H. & Rubin, M. H. Experimental evaluation of a two-photon wave packet in type-II parametric downconversion. *J. Opt. Soc. Am. B* **12**, 859-862 (1995).

31. Ou, Z. Y. & Mandel, L. Observation of spatial quantum beating with separated photodetectors. *Phys. Rev. Lett*. **61**, 54–57 (1988).

32. Kim, H., Lee, H. J., Lee, S. M. & Moon, H. S. Highly efficient source for frequency-entangled photon pairs generated in a 3rd-order periodically poled MgO-doped stoichiometric $LiTaO_3$ crystal. *Opt. Lett*. **40**, 3061-3064 (2015).

## Acknowledgements

This work was supported by the Basic Science Research Program through the National Research Foundation of Korea funded by the Ministry of Education, Science and Technology (No. 2015R1A2A1A05001819 and No. 2016R1D1A1B03936222). Also, this work was supported by the Measurement Research Center (MRC) Program for Korea Research Institute of Standards and Science. S.M.L. acknowledges the support of the National Research Foundation of Korea (NRF) grant (No. 2014R1A1A2054719), the R&D Convergence Program of the NST (No. CAP-15-08-KRISS), and the KRISS project 'Convergent Science and Technology for Measurement at the Nanoscale' of Republic of Korea.

## Author Contributions

H.K. and H.S.M. conceived the project. H. K. designed the experimental setup and performed the experiment. H.K., S.M.L., O.K. and H.S.M. discussed the results and contributed to writing the manuscript.

## Additional Information

**Competing financial interests:** The authors declare no competing financial interests.